\documentclass{elsart}

\usepackage{epsfig}

\usepackage{amssymb}

\begin{document}

\begin{frontmatter}


% use the thanksref command within \title, \author or \address for footnotes:
% \author{Name\thanksref{label2}}
% including your email address:
% \address{Address\thanksref{email}}
% \thanks[email]{E-mail: }

\title{Improved limits on photon velocity oscillations}


\author{A.~De~Angelis}

\address{
Dipartimento di Fisica dell'Universit\`a di Udine and INFN Trieste, Italy}

\author{R. Pain}

\address{LPNHE, Universit\'es Paris VI \& VII and IN2P3/CNRS, Paris, France}

\begin{abstract}
% Text of abstract
The mixing of the photon with a hypothetical sterile paraphotonic state 
would have consequences on the cosmological propagation of photons. 
The absence of distortions in the optical spectrum of 
distant Type Ia supernov\ae\, 
allows to extend by two orders of magnitude the 
previous limit on the Lorentz-violating 
parameter~$\delta$ associated to the photon-paraphoton transition, 
extracted from the abscence of distortions in the spectrum of the 
cosmic microwave background. 
The new limit
is consistent with the interpretation of the dimming of  distant Type Ia 
supernov\ae\,
as a consequence of a nonzero cosmological constant. 
Observations of gamma-rays from active galactic nuclei allow to further
extend the limit on~$\delta$ by ten orders of magnitude. 
\end{abstract}

\end{frontmatter}

% main text
%\section{}
%\label{}

%\begin{thebibliography}{00}

%\bibitem{}

%\end{thebibliography}

%\end{document}

%==================> Note text          =====> To be filled <======%
%\section{Introduction}

\baselineskip 16pt

The existence of a second photon (paraphoton) mixing to the ordinary one 
was first postulated in \cite{geo} to explain a presumed anomaly in the 
spectrum of the Cosmic Microwave Background (CMB). 
In that model, the anomaly was 
attributed to a mass mixing of the two photons analogous to the
oscillation of neutrinos. 
An ordinary photon oscillates with the time $t$ in such a way that its 
probability
to stay as such can be written as
\begin{equation}
P(t) = 1-\sin^2(2\phi)\,\sin^2{(\rho \mu^2 \, t/\omega )} \, , \label{oldmo}  
\end{equation}
where $\omega$ is the frequency of the photon,
$\rho = c^2/4\hbar^2$, $\phi$ is the mixing angle and
$\mu$ is the mass difference between the two mass eigenstates
(i.e., the mass of the additional photon if the standard one is massless).
Thus the oscillation probability decreases with the increasing photon energy.
The thermal nature of the CMB has then been established by 
COBE~\cite{cobe} and the anomaly has vanished;
from the agreement of the CMB with the blackbody radiation,
a second photon with mass $\mu \neq 0$ 
maximally mixing to the standard one has been
excluded \cite{debesmo} at the level of
\begin{equation}
\mu < 10^{-18} {\rm{eV}} \, ,
\end{equation}
to be compared with the present limit of $m_\gamma < 2 \times 10^{-16}$ eV 
on the photon mass \cite{pdg}. 
Eq.~(\ref{oldmo}) shows that, in this kind of model, 
one achieves maximum sensitivity to the mixing by studying
low-frequency radiation. 

A different model \cite{glas} has been recently motivated
by the possible existence of tiny departures from
Lorentz invariance \cite{cole}, which could explain the presence of
cosmic rays beyond the Greisen-Zatsepin-Kuzmin (GZK) 
cutoff \cite{gzk}.
An additional photon state would
experience Lorentz non-invariant  
mixing with the standard 
one, and the two eigenstates
would propagate in any direction at slightly different velocities, say, 
$c$ and $(1+\delta)\,c$.
Velocity oscillations of
photons could also result from violations of the equivalence principle in a
Lorentz invariant theory \cite{gasp}, or from the mixing with
photons in a ``shadow'' universe~\cite{sak}.

The paraphoton in \cite{glas} is sterile;
photons emitted by ordinary matter 
evolve in such a way that the non-interacting component develops with time,
and the probability for an ordinary photon to stay as such 
oscillates with time according to:
\begin{equation}
P(t) = 1-b^2\,\sin^2{(\delta \,\omega t/2)} \label{eosc} 
\end{equation}
with $\omega$ the frequency of the detected photon and 
$b^2\equiv \sin^2{(2\phi)}$, where $\phi$ is the mixing angle. 
We are concerned with the large mixing ($b\sim 1$) and small $\delta$
domain. 

\begin{figure}
\begin{center}
\epsfig{file=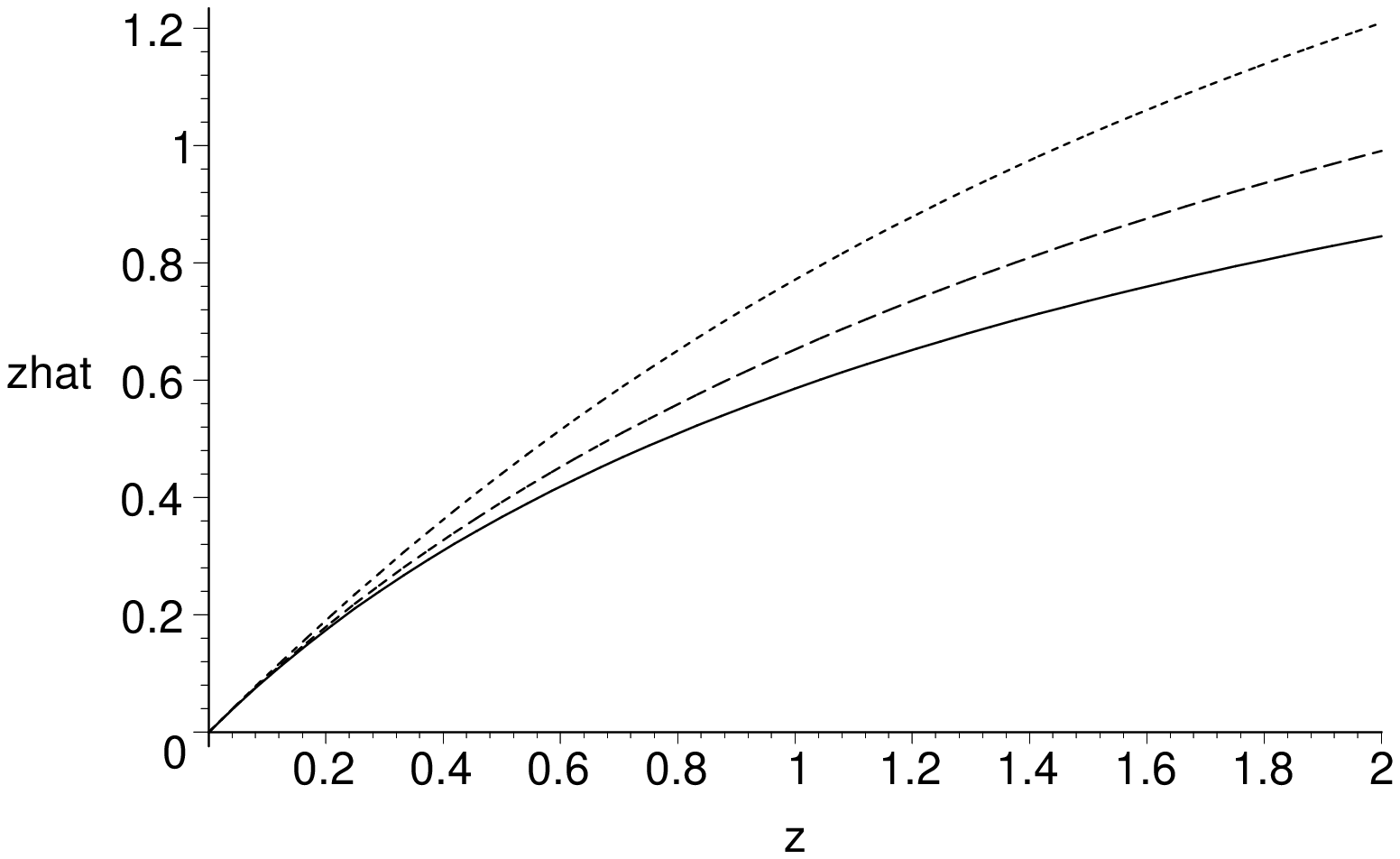,width=\linewidth}
\end{center}
\caption{The function $\hat{z}(z)$: for 
$\Omega = 1$ and $\Omega_\Lambda = 0$ (solid line); 
for $\Omega = 1$ and $\Omega_\Lambda = 0.7$ (dashed line);
and for $\Omega = 0.3$ and $\Omega_\Lambda = 0$ (dotted line).
}
\label{fig1}
\end{figure}

The extinction coefficient on light from a source at redshift $z$,
due to velocity oscillations, 
can be written as a function of $z$ as:
\begin{equation} 
P(z) =1-b^2\sin^2( \delta\,\omega \,\hat{z}/2H_0) \, , \label{eoscc} 
\end{equation}
where
\begin{equation}
\hat{z} = H_0 \int_0^{z} (1+\zeta)\,d\zeta\,(dt/d\zeta)\;, \label{erep} 
\end{equation}
$H_0=h \times 100\ {\rm  km/s}\cdot$Mpc, and the redshift-time relation
can be written:
\begin{eqnarray*}
\frac{dy}{y} & = & -H(y)\,dt=-H_0\,\left[(1-\Omega)y^2 +(\Omega-\Omega_\Lambda)y^3
+\Omega_\Lambda\right]^{1/2 }\,dt %& \\ & \Longrightarrow & \hat{z}(z) = ??
\end{eqnarray*}
with $y=(1+\zeta)$, 
$\Omega_\Lambda$ the cosmological constant and $\Omega_{\rm M} =
\Omega-\Omega_\Lambda$.
The function $\hat{z}(z)$ is plotted in Figure \ref{fig1} for 
$\Omega = 1$ and $\Omega_\Lambda = 0$; for $\Omega = 1$ and $\Omega_\Lambda = 0.7$;
and for $\Omega = 0.3$ and $\Omega_\Lambda = 0$ respectively.

The analysis of Ref. \cite{glas} improves the previous limit \cite{old} by ten
orders of magnitude by studying the departures of the CMB
from the thermal spectrum; a limit ${b\delta}/{h} < 1.6 \times 10^{-32}$ 
is obtained, which corresponds to
\begin{equation}
\delta < 1.1 \times 10^{-32} \label{lglas} 
\end{equation}
for $h=0.7$ and $b=1$.

By inspecting Eq. (\ref{eosc}), one can see
that, in this kind of model, 
one achieves maximum sensitivity to the mixing by studying
high energy radiation. 

The author of Ref. \cite{glas} observes that velocity oscillations might
systematically distort the spectra of more distant sources 
and change their apparent magnitudes, complicating the determination of
cosmological parameters via the measurements of redshifts and
apparent magnitudes of distant Type~Ia supernov\ae\, (SNe~Ia) 
\cite{SCP}\cite{HZT}. 
The effect could be such that data become consistent with a
``standard'' $\Omega=1$ and $\Omega_\Lambda =0$ universe. 

The presence of ``dark energy'' has been recently independently 
confirmed \cite{deber,unce}
from a combination of measurements of the CMB and the distribution 
of galaxies on large scales.
However, the uncertainties in measuring the cosmological
parameters $\Omega_{\rm M}$ and $\Omega_\Lambda$ do not allow to rule out
the hypothesis that part of the dimming is due to
photon mixing \cite{ariel}. Given the current uncertainties of $\pm0.1$ in
the measurement of $\Omega_{\rm M}$ \cite{unce} 
and assuming that the Universe is flat
with $\Omega_{\rm M}=0.3$, we can compute an upper limit for the contribution
to the dimming of SNe~Ia coming from photon-paraphoton oscillations. 

\begin{figure}
\begin{center}
\epsfig{file=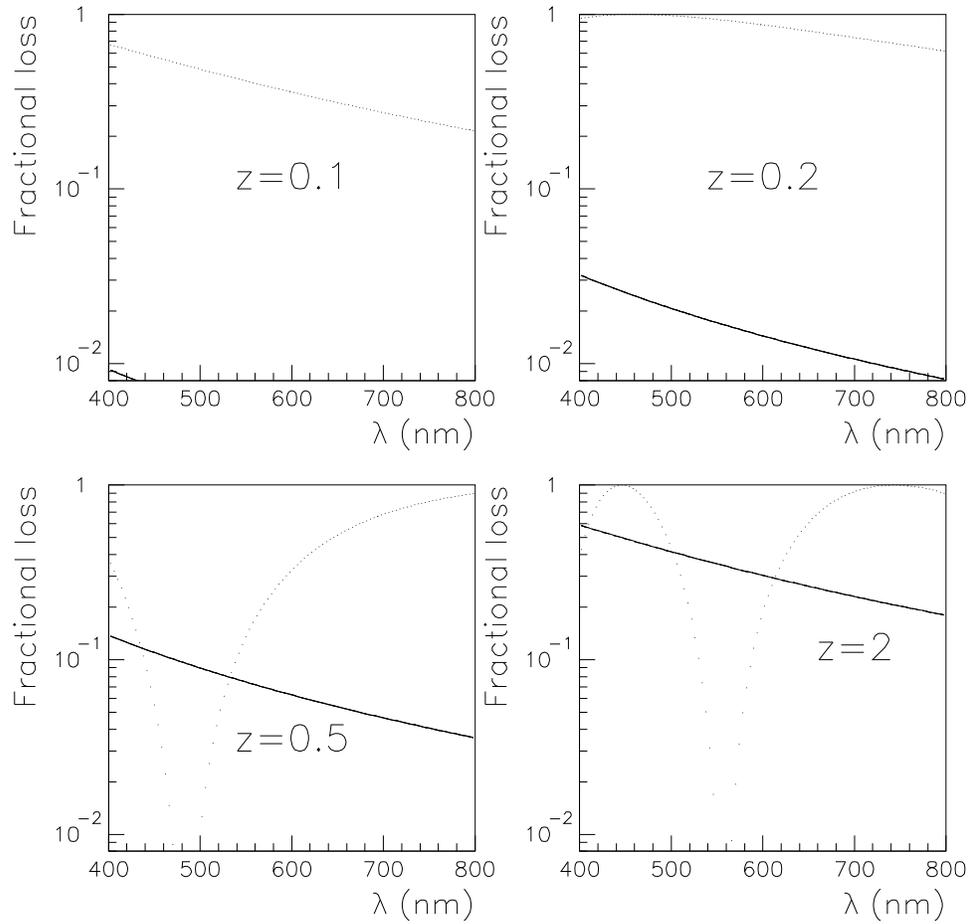,width=\linewidth}
\end{center}
\caption{Fractional loss of photons due to the oscillation into paraphotons 
as a function of the wavelength near the optical region. 
The curves for $\delta=10^{-32}$ (dotted) 
and $\delta=10^{-33}$ (solid) are shown.}  
%in the plots corresponding to $z=2$ the curves 
%for $\delta=10^{-33}$ (the curve displaying oscillations), $\delta=10^{-34}$,
%and $\delta=10^{-35}$ (lowest curve) are shown.}
%In the plots corresponding to $z=0.1$, $z=0.2$ and $z=0.5$ the curves 
%for $\delta=10^{-33}$ (upper curve) and $\delta=10^{-34}$ are shown;  
%in the plots corresponding to $z=2$ the curves 
%for $\delta=10^{-33}$ (the curve displaying oscillations), $\delta=10^{-34}$,
%and $\delta=10^{-35}$ (lowest curve) are shown.}
\label{fig2}
\end{figure}

The fractional loss of photons due to the oscillation into paraphotons 
is shown in Figure \ref{fig2}
as a function of the wavelength in the optical region for different
redshifts of the source. The curves were computed
for $\delta=10^{-32}$ (dotted lines) 
and $\delta=10^{-33}$ (solid lines).
A maximal mixing with $\delta = 10^{-32}$,
of the order of the limit (\ref{lglas}), would severely distort the
optical spectrum for a redshift $z=0.5$, and can thus be excluded. A value 
$\delta \sim 10^{-33}$ would fake a large part of the effect observed 
in \cite{SCP}\cite{HZT}.
The function $\hat{z}(z)$ used in Figure~\ref{fig2} is 
calculated assuming $\Omega=1$ and $\Omega_\Lambda=0$; as seen
in Figure~\ref{fig1}, this is a conservative hypothesis
compared to assuming, e.g., $\Omega=1$ and $\Omega_\Lambda=0.7$.

\begin{figure}
\begin{center}
\epsfig{file=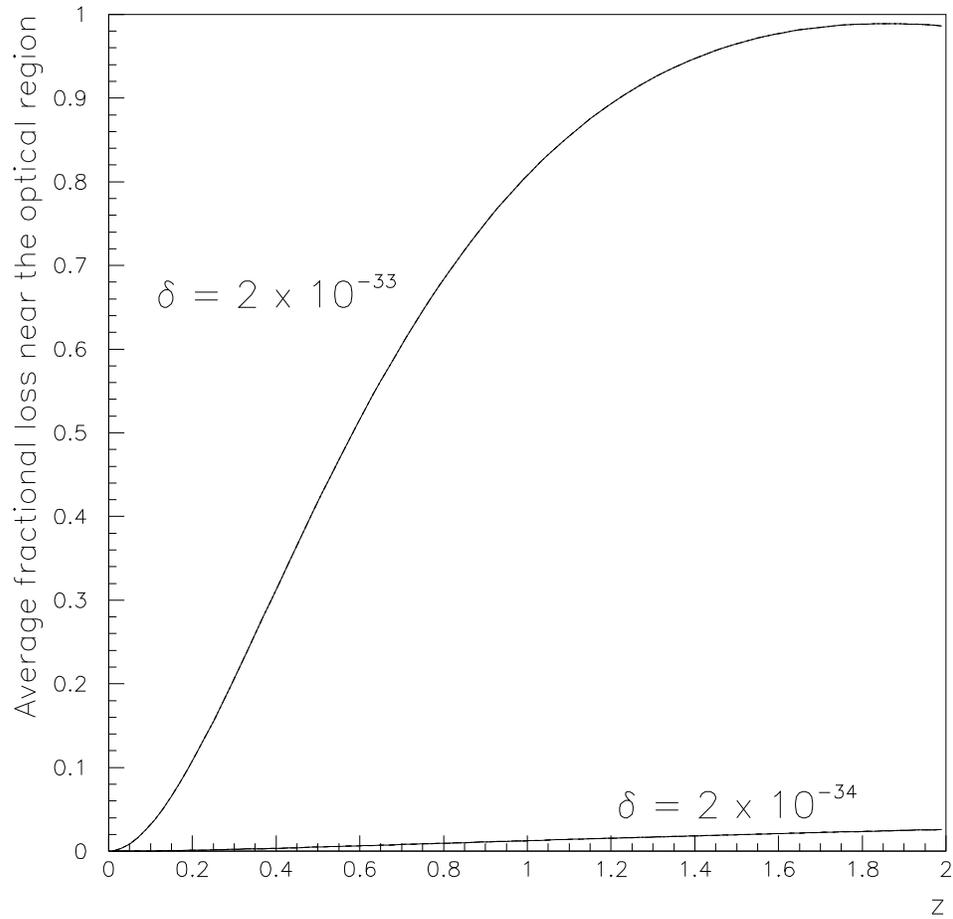,width=\linewidth}
\end{center}
\caption{Fractional loss of photons emitted in the $B$-band (see text) 
due to the oscillation into paraphotons, 
as a function of the redshift $z$. The two curves correspond to   
$\delta=2 \times 10^{-33}$ (upper curve) and $2 \times \delta=10^{-34}$ (lower curve).}
\label{fig3}
\end{figure}

The improved sensitivity coming from the dimming of distant 
sources is made evident 
in Figure~\ref{fig3}, where the fractional 
loss of photons emitted in the 
$B$-band\footnote{The standard $B$-band goes 
from $\lambda=380$ nm to $\lambda=520$ nm (see~\cite{bessel}).}
due to the oscillation into paraphotons is displayed.
We computed that the modulation would affect the
observation of supernov\ae\ at $z=0.5$ in the rest-frame 
$B$-band, while consistent with the current $\Omega_\Lambda$ value,
for 
\begin{equation}
\delta < 7 \times 10^{-34} \, ,
\end{equation}
one order of magnitude smaller than the limit (\ref{lglas}). 

An independent limit can be obtained from the absence of distortions 
in the spectrum of SNe~Ia between low-$z$ and high-$z$.
The flux $F$ of SNe~Ia is 
often also measured in a second band pass, usually the 
$V$-band\footnote{The standard $V$-band goes from 
$\lambda=490$ nm to $\lambda=680$ nm (see~\cite{bessel}).},
in order to address extinction effects. Using data of \cite{SCP}, and 
assuming the the average extinction of distant and nearby SNe~Ia is 
identical, we can compute the limit coming from the absence of 
distortions in the spectra from distant SNe spectra. 

A limit of 
$$\frac{(F(B)/F(V))_{z\sim0.5}}
       {(F(B)/F(V))_{z\sim 0}} > 0.980 \; {(95\% \, C.L.)}$$
is obtained from the data of Ref. \cite{SCP}; from this one obtains:
\begin{equation}
\delta < 2 \times 10^{-34} \, .
\end{equation} 
More constraining limits could be 
reached if no distortions were observed in the
spectrum of more distant supernov\ae.

\begin{figure}
\begin{center}
\epsfig{file=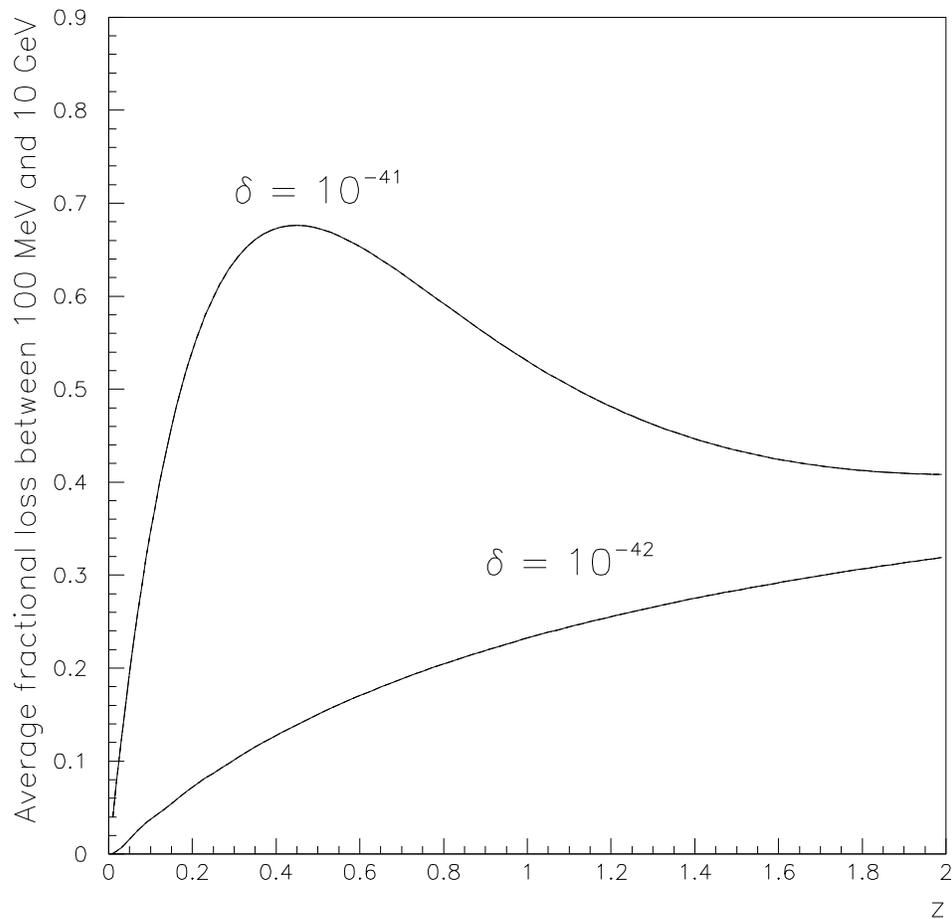,width=\linewidth}
\end{center}
\caption{Fractional loss of photons due to the oscillation into paraphotons 
in the $\gamma$-ray region, averaged over the frequency between 
$E=100$ MeV and $E=10$ GeV 
(assuming an energy spectrum proportional to $E^{-2}$), 
as a function of the redshift $z$. The two curves correspond to     
$\delta=10^{-41}$ (upper curve) and $\delta=10^{-42}$ (lower curve).}
\label{fig4}
\end{figure}

The presence of the term $\omega$ in Eq. (\ref{eoscc}) is such that the
sensitivity to $\delta$ improves further 
by making observations in the $\gamma$-ray region.
In Figure \ref{fig4}
the fractional loss of photons due to the oscillation into paraphotons 
in the $\gamma$-ray region is averaged over the frequency between 
$E=100$ MeV and $E=10$ GeV, for a radiation with a
power-law spectrum $E^{-2}$. 
The non-observation of distortions in the $\gamma$-ray spectrum at  
$z \sim 1$, in an energy region corresponding to
the sensitivity of GLAST \cite{glast},
can thus allow to set limits between $\delta < 10^{-41}$
and  $\delta < 10^{-42}$.

A rule-of-thumb relation on the value of $\delta$ which could have sizable
effects on the propagation from a given redshift $z$ 
of a photon of energy $E$ can be obtained by
setting to unity the argument of the sine in  Eq. (\ref{eoscc}):
\begin{equation}
\delta \sim  
\frac{3 \times 10^{-33}}{2(1-1/\sqrt{1+z})}  \, 
\left( \frac{\mathrm{1\,eV}}{E} \right) \, . \label{rot}
\end{equation}
The high-energy gamma data from Mkr 501 \cite{ah99}
at a redshift $z \simeq 0.034$, if interpreted 
as in agreement with models \cite{ko99}, allow to set
a model-dependent limit
\begin{equation}
\delta < 10^{-44} 
\end{equation} 
from Eq. (\ref{rot}) assuming that there is no distortion at an energy 
$E \sim 10$ TeV.

\vspace*{1cm}

{\bf{Acknowledgments -}} Many thanks to 
Pierre Astier, Etienne Barrelet, Said Bouaissi,
Francesco Longo and Oriana Mansutti. 
One of us (A.D.A.)
would like to thank the Universit\'e Pierre et Marie Curie - Paris VI
for the kind hospitality.

\newpage

\newpage

\newpage

\newpage

\end{document}